\documentclass{article}

\usepackage[preprint,nonatbib]{neurips_2021}

\usepackage[utf8]{inputenc} 
\usepackage[T1]{fontenc}    
\usepackage{hyperref}       
\usepackage{url}            
\usepackage{booktabs}       
\usepackage{amsfonts}       
\usepackage{nicefrac}       
\usepackage{microtype}      
\usepackage{xcolor}         
\usepackage{subcaption}
\usepackage{graphicx}
\usepackage{algorithm, algorithmic}
\usepackage{xcolor}
\usepackage{amsmath}

\title{Piano Fingering with Reinforcement Learning}

\author{%
  Pedro Ramoneda \\
  Music Technology Group\\
  Pompeu Fabra University\\
  Barcelona, 08018 \\
  \texttt{pedro.ramoneda@upf.edu}\\
  \And
  Marius Miron \\
  Music Technology Group\\
  Pompeu Fabra University\\
  Barcelona, 08018 \\
  \texttt{marius.miron@upf.edu}\\
  \And
  Xavier Serra \\
  Music Technology Group\\
  Pompeu Fabra University\\
  Barcelona, 08018 \\
  \texttt{xavier.serra@upf.edu}\\
  
}

\begin{document}

\maketitle

\begin{abstract}
  Hand and finger movements are a mainstay of piano technique. Automatic Fingering from symbolic music data allows us to simulate finger and hand movements. Previous proposals achieve automatic piano fingering based on knowledge-driven or data-driven techniques. We combine both approaches with deep reinforcement learning techniques to derive piano fingering. Finally, we explore how to incorporate past experience into reinforcement learning-based piano fingering in further work.
\end{abstract}

\section{Introduction}





Piano technique is considered a fundamental performance skill~\cite{neuhaus,chiantore,lech,nieto,lev}. Through fingering the notes of a score, we can model the technique of hands and fingers piano movements~\cite{Ramoneda22,RamonedaTFM}. Rather than a fixed set of rules, piano fingering is a creative and flexible process individualised for each pianist~\cite{chiantore}.

Piano fingering plays an important role in the realization of the music performance~\cite{neuhaus}. To that extent, pianists must adapt the fingering at each moment according to the subsequent fingering patterns’ needs~\cite{nieto}. Fingering must preserve the musical content of the work in all its facets: articulation, tempo, dynamics, rhythm, style and character~\cite{nieto}. On the other hand, it has to be as comfortable as possible~\cite{lech}. 
Moreover, piano fingering changes individually according to the size of the hand~\cite{lev}.

In this paper we model the problem of automatic fingering by using reinforcement learning (RL). The code is available online~\cite{code1}.
The RL policy is defined as such to reduce the hand's movement. To that extent, the reward is higher if there are fewer hand positions. 
Besides, the possible finger combinations are very large when fingering a score. However, the optimal combinations, which are the most comfortable while respecting the musical content, are more limited. 

The direct application of our proposed method is to give feedback to the piano students to improve their fingering. We aim at presenting various alternative finger's combinations to the music student. Our RL method may offer different solutions corresponding to different iterations and to different fingers combinations. These solutions may help the music students improve their technique.

Several proposals aim at modeling piano fingering with various techniques, from expert systems~\cite{parn1,parn2} through local search algorithms~\cite{VNS,PP} to data-driven methods~\cite{N1, N2}. In contrast, we aim at codifying the expert knowledge on the reward function of a RL algorithm. Moreover, we aim to optimising to a broader term, like the local search algorithms\cite{VNS,PP}, thanks to the RL sparsity property. Besides, our proposed method seeks to optimize from note to note each action, the Markov decision process, as data-driven proposals~\cite{N1, N2}.

The remainder of this paper is structured as follows. We present the RL fingering method in Section~\ref{baseline}. We expose the preliminary results and the further work in Section~\ref{results} and Section~\ref{work}.

\section{Methodology}
\label{baseline}

In the present approach, an agent interacts with a score understood as the environment. Consequently, each score is a different environment. The environment is reduced to only the right hand because we can replicate the left hand by symmetry~\cite{VNS,N1}. We define the state $s$ associated with the finger a tuple comprising $(cf, cn, nn)$ being $ch$ the current finger, $cn$ the current note and $nn$ the next note, and with the first finger of the sheet known. Note, the size of the notes $cn$, $nn$ encoding space is determined by the melodic range.

The action $a$ is the next finger according to its policy $\pi(a|s)$. We define a set of fingering rules encoded in the reward function $r(s|a)$ dependent on the finger selected as action $a$. The reward function $r$ gives a positive reward if no hand position changes and the negative reward is the opposite. In addition, $r$ negatively rewards the anatomically not feasible actions. Finally, the Q function of a policy $\pi$ is estimated with a Fully Connected-based DQN. The complete RL algorithm scheme can be found in Figure~\ref{alg:DQN}.

\section{Evaluation}
\label{results}

We conduct five experiments to test the behaviour of the fingering algorithm EX1, EX2, EX3, EX4 and EX5. EX1 contains a sequence of notes with the same pitch and rhythmic figure. This experiment aims to test whether the RL agent learns to use the same finger in every note. In EX2, we have a partial split scale of five ascending notes and the same five descending notes. In this experiment, we test whether the RL agent learns not to change the hand position. The EX3 is a piece of music that does not change the hand position throughout its length. Similarly to EX2, EX3 aims at keeping the same position but in a complex environment. EX is a C major scale. Therefore, the RL agent should learn to perform only two hand position changes. The fifth test is a piece with the melodic range of the C major scale. In this case, we want to test whether the RL agent learns to keep the same two hand position as EX4 but in a complex environment. All these experiments have been carried out with various improvements and with different numbers of episodes.

\begin{figure}[h!]
\centering

\begin{subfigure}[b]{0.86\textwidth}
   \includegraphics[width=1\linewidth]{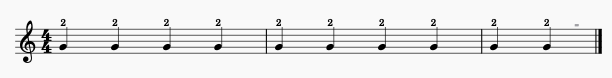}
   \caption{}
\end{subfigure}

\begin{subfigure}[b]{0.86\textwidth}
   \includegraphics[width=1\linewidth]{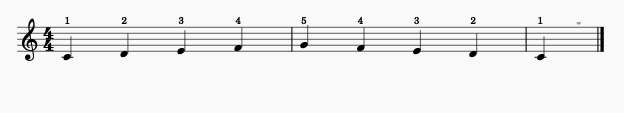}
   \caption{}
\end{subfigure}

\caption{(a) Test 1 trained on 1000 episodes. (b) test 2 trained on 100}
\label{comparison12}
\end{figure}

\begin{figure}[h!]
\centering

\begin{subfigure}[b]{0.92\textwidth}
   \includegraphics[width=1\linewidth]{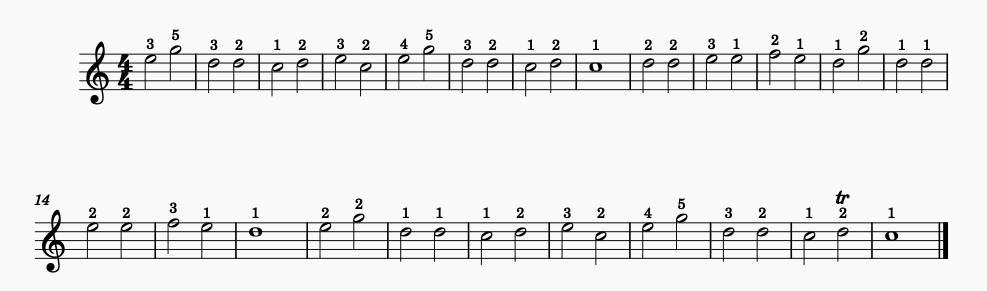}
   \caption{}
\end{subfigure}

\begin{subfigure}[b]{0.92\textwidth}
   \includegraphics[width=1\linewidth]{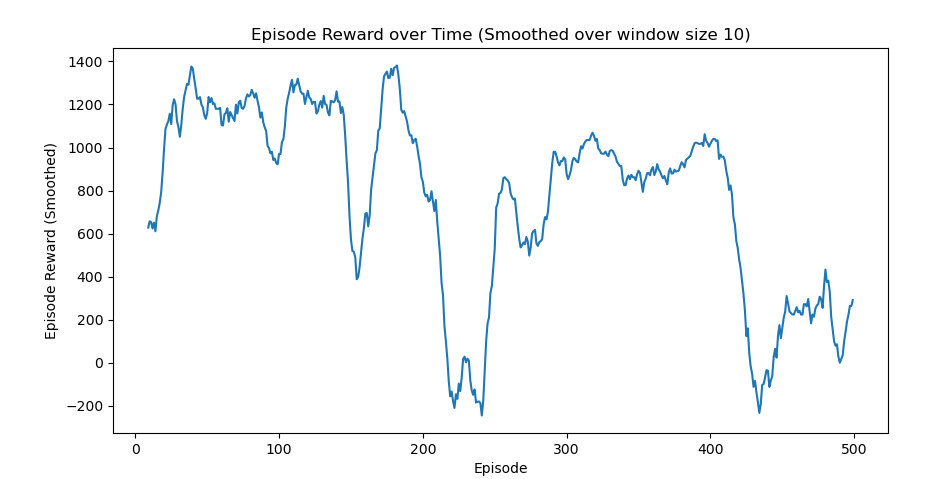}
   \caption{}
\end{subfigure}

\caption{(a) Test 3 trained on 200 episodes. (b) Episode/reward over time.}
\label{test3}
\end{figure}

\begin{figure}[h!]
\centering

\begin{subfigure}[b]{0.92\textwidth}
   \includegraphics[width=1\linewidth]{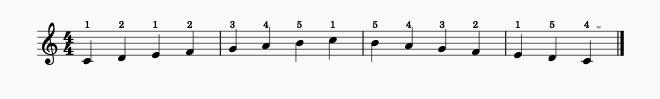}
   \caption{}
\end{subfigure}

\begin{subfigure}[b]{0.92\textwidth}
   \includegraphics[width=1\linewidth]{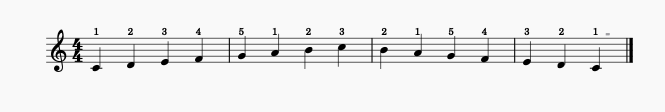}
   \caption{}
\end{subfigure}

\begin{subfigure}[b]{0.92\textwidth}
   \includegraphics[width=1\linewidth]{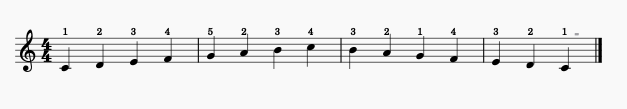}
   \caption{}
\end{subfigure}

\begin{subfigure}[b]{0.92\textwidth}
   \includegraphics[width=1\linewidth]{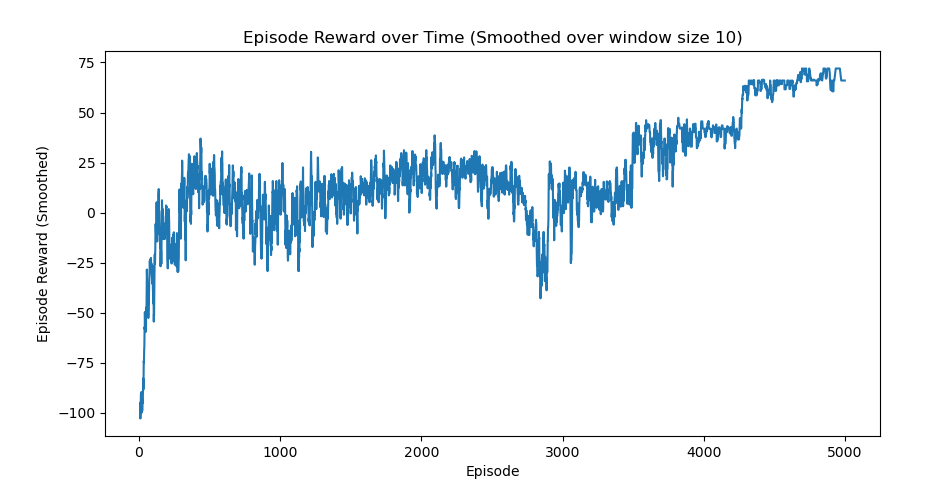}
   \caption{}
\end{subfigure}

\caption{(a) Test 4 trained on 500 episodes. (b) Test 4 trained on 2000 episodes. (c) Test 4 trained on 5000 episodes. (d) Evolution of reward in 5000 episodes }
\label{test4}
\end{figure}

\begin{figure}[h!]
\centering

\begin{subfigure}[b]{0.92\textwidth}
   \includegraphics[width=1\linewidth]{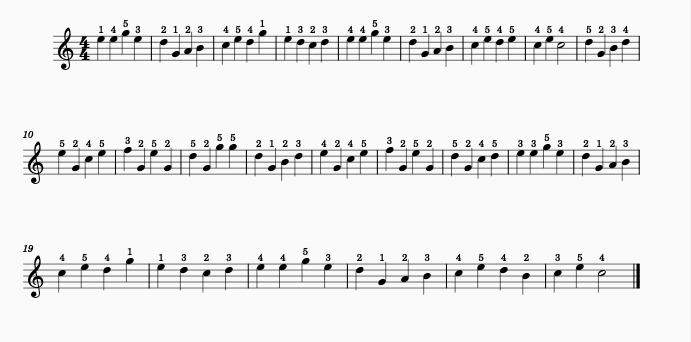}
   \caption{}
\end{subfigure}

\begin{subfigure}[b]{0.92\textwidth}
   \includegraphics[width=1\linewidth]{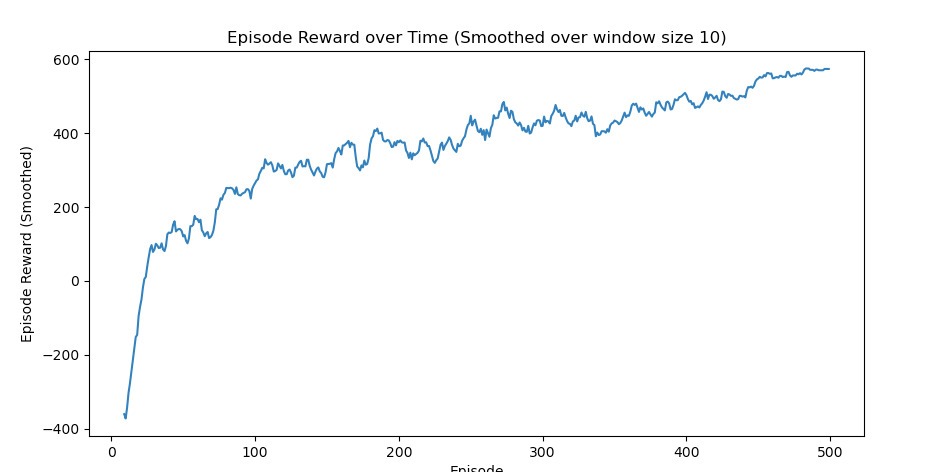}
   \caption{}
\end{subfigure}

\caption{(a) Test 5 trained on 500 episodes. (b) Episode/reward over time.}
\label{test5}
\end{figure}

For the first two experiments, due to their simplicity, the results are as expected. In EX1, the same finger is playing all the notes sequence, and in the second experiment, there is no change of hand position. In the EX1, each note is encoded as 88 grooves, while rest of the experiments, the encoding is reduced to the melodic range, improving the convergence time, as is shown in Figure~\ref{comparison12}. In some cases, the reward function oscillates when following a less conservative strategy and trying to explore for too long, as exposed in EX3, Figure~\ref{test3}. The evolution of reinforcement learning can be seen in EX4, shown in Figure~\ref{test4}. The trial and error to achieve optimum fingering resemble the process carried out by a pianist. In EX5, Figure~\ref{test5}, we can see the reward function evolves little by little, approaching the expected result. Although the preliminary reinforcement results surpass our expectations, the system requires more time to achieve the desired results compared with offline methods. In Section~\ref{work}, we plan strategies to solve the convergence issue.

\section{Conclusions and future work}
\label{work}

\begin{figure}[h!]
 \centering
 \includegraphics[width=160mm]{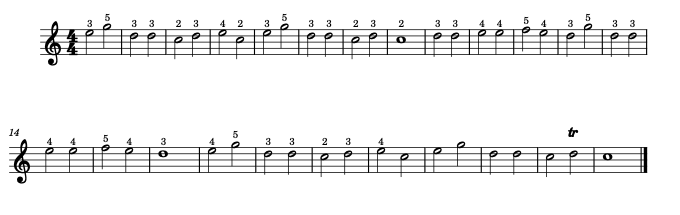}
  \caption{Test 3 synthetic model results.}
 \label{synthetic}
\end{figure}

Although pianists learn to finger throughout their career by trial and error, they do not start from zero in every score/environment. All the works they have played and also fingered before help musicians to finger a new score. RL previous approaches attempt to address this through multi-agent RL~\cite{multi},  bringing offline and online RL~\cite{offline} or fine-tuning the models with RL~\cite{magentaRL}. We have chosen Pianoplayer~\cite{PP}, a non-data-driven algorithm that summarises the concept of comfortable, to create a synthetic dataset of more than 1500 piano scores. This dataset has been created by cross-referencing Musescore public domain works with the most famous classical piano composers. Thereby we want to demonstrate that it is possible to incorporate synthetic knowledge in a supervised way with a shallow GRU network architecture. This architecture was previously proposed for time series RL~\cite{ts}. The results also surpass our expectations, and the implementation is available online~\cite{code2}. The shallow architecture can imitate expert systems in a 78\% balanced accuracy and feasible combinations of data as shown in Figure~\ref{synthetic} data. Departing from the presented synthetic mode, in the future, we will explore different ways of incorporating the existing fingering knowledge into reinforcement learning methods.
 
\section*{Acknowledgement}

I would like to thank my two colleagues Sergio Izquierdo and Julia Guerrero from the University of Freiburg/University of Zaragoza, for all the help in understanding the RL paradigm and the discussion about this project.

\medskip

{

}

\begin{algorithm}[h]
    \caption{Deep Q-Learning with Experience Replay}
    \label{alg:DQN}
    \begin{algorithmic}
	\STATE Initialize replay memory $\mathcal{D}$ to capacity $N$
	\STATE Initialize action-value function $Q$ with two random sets of weights $\theta, \theta'$
	\FOR{$episode = 1,M$}
	    \FOR{$t = 1,T$}
		\STATE Select a random action $a_t$ with probability $\varepsilon$
		\STATE Otherwise, select $a_t = {\arg\max}_a Q(s_t, a; \theta)$
		\STATE Execute action $a_t$, collect reward $r_{t+1}$ and observe next state $s_{t+1}$
		\STATE Store the transition $(s_t, a_t, r_{t+1}, s_{t+1})$ in $\mathcal{D}$
		\STATE Sample mini-batch of transitions $(s_j, a_j, r_{j+1}, s_{j+1})$ from $\mathcal{D}$
		
		\IF{$s_{j+1} \text{is terminal}$}
		\STATE $y_j = r_{j+1}$
		
		\ELSE
		\STATE $y_j = r_{j+1} + \gamma \max_{a'} Q(s_{j+1}, a'; \theta')$ 
		\ENDIF

		\STATE Perform a gradient descent step using targets $y_j$ with respect to the online parameters $\theta$
		\STATE Every $C$ steps, set $\theta' \leftarrow \theta$
	    \ENDFOR
	\ENDFOR
    \end{algorithmic}
\end{algorithm}

\begin{figure}[h!]
 \centering
 \includegraphics[width=160mm]{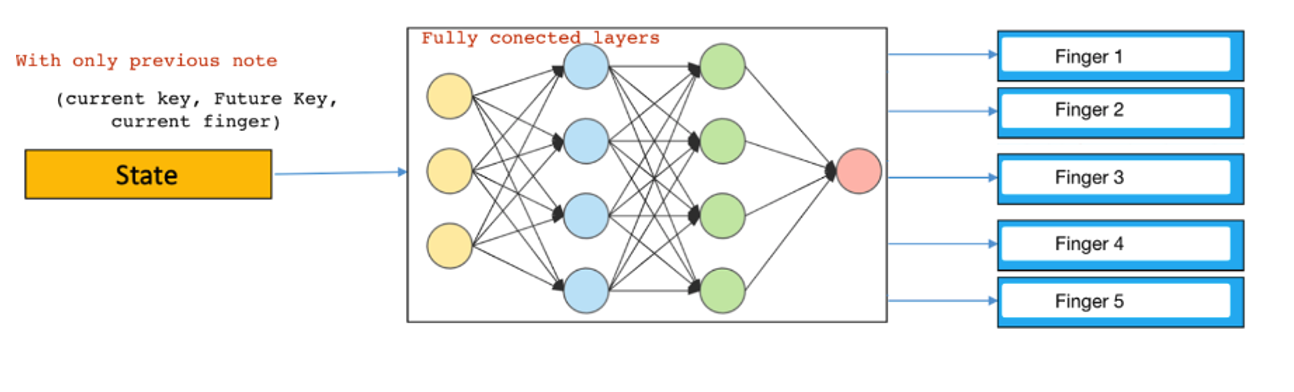}
  \caption{Deep Q neural network diagram.}
 \label{diagram_reinforcement}
\end{figure}

\end{document}